# Factorial Representations of Compact Lie Groups, Wigner Sets and Locally Invariant Quantum Fields

J Moffat and C Wang

*Dept of Physics, University of Aberdeen, UK*

**Abstract**

The fibre bundle construct defined in our previous work continues to be the context for this paper; quantum fields composed of fibre algebras become liftings of; or sections through; a fibre bundle with base space a subset of curved space-time. We consider a compact Lie group such as *SU(n)* acting as a local gauge group of automorphisms of each fibre algebra **A(x)**. Compact Lie groups, represented as gauge groups acting locally on quantum fields, are key elements in electroweak and strong force unification. In our recent joint work we have focused on the translational subgroup of the Poincare group as the generator of local diffeomorphism invariant quantum states. Here we extend those algebraic non-perturbative approaches to address the other half of unification by considering the existence of quantum states of the fibre algebra **A(x)** invariant to the action of compact non-abelian Lie groups. Wigner sets are complementary to little groups and we prove they have the finite intersection property. Exploiting this then allows us to show that invariant states are common in the sense that the weakly closed convex hull of every normal (density matrix) state contains such an invariant state. From these results and our related research emerges the existence of a locally invariant density matrix quantum state of the field *A=A(x)* and a locally diffeomorphism invariant 'gravity state' of the field compatible with a massless supersymmetric graviton.

**Introduction and Context**

We consider a compact Lie group $G$ acting as a gauge group of automorphisms of the fibre algebra **A(x)**. Recall [1-4] that **A(x)** is a Quantum Operator Algebra; a von Neumann algebra with trivial centre acting (up to isomorphism) on a separable Hilbert space *F(x)*, and that locally the fibre bundle is a product bundle. We define a section algebra *O(D)* as the closure in the ultraweak operator topology of the set of all such fibre algebras with the algebraic operations defined fibrewise $O(D) = \{\mathbf{A}(\mathbf{x}); \mathbf{x} \in D\}^{-\sigma w}$ with the base $D$ a suitable subset of spacetime of physical interest. It can be characterised as the dual space of it predual $O(D)_*$ which consists of all normal, density matrix, states. Consider a representation of the compact Lie group $G$ as automorphisms $\alpha : g \to \alpha_g(A)$ for $A \in O(D), g \in G$. In [3] we introduced the crossed product algebra $O(D) \times G$ and the linkage between partitions and von Neumann entropy (also known as quantum or entanglement entropy).

Assume (by taking a faithful representation if necessary) that $O(D)$ acts on a Hilbert space $H$. Define the Dirac distribution function $\varepsilon_g$ to take the value 1 at g and zero elsewhere on G. Then $\{\varepsilon_g ; g \in G\}$ is an orthonormal basis for the Hilbert space $l^2(G)$. Given $l^2(G)$ and $H$ we can form the tensor product Hilbert space $H \otimes l^2(G)$. If $\alpha : g \to \alpha_g(A)$ for $A \in O(D), g \in G$, define;

$$U_h(x \otimes \varepsilon_g) = x \otimes \varepsilon_{gh^{-1}} \text{ for } x \in H, \ g, h \in G$$
$$\Phi(A)(x \otimes \varepsilon_g) = \alpha_g(A)x \otimes \varepsilon_g \text{ for } A \in O(D), g \in G$$

Then $U_h$ extends to a unitary operator on $H \otimes l^2(G)$ and the mapping $h \to U_h$ is a unitary representation of the compact Lie group $G$ on $H \otimes l^2(G)$. Similarly, $\Phi(A)$ extends to a bounded linear operator on $H \otimes l^2(G)$ for all sections $A$ in $O(D)$ and the unitary representation $h \to U_h$ implements the automorphic representation $h \to \alpha_h$.

The transformation $\Phi$ is an ultraweakly continuous isomorphism of $O(D)$ and it follows that $\Phi(O(D))$ is a von Neumann algebra embedding. Finite sums $\sum_j U_{g_j} \Phi(A_j)$ form a *algebra which contains $\Phi(O(D))$. The cross product algebra $O(D) \times G$ is defined as its closure for the ultraweak operator topology (equivalently its double commutant).

Given a partition of a locally curved space-time region into subsets $D(1), D(2),....,D(n)$ we have from early foundational work [5] that;

$(O(D1) \times G) \otimes O(D2) \times G)$ is isomorphic to $(OD1 \otimes OD2) \times (G \times G = G^2)$. If we assume that by induction we have a similar result for the case *n-1* then;

Applying [5] to the algebras $(\bigotimes_{j=1}^{n-1}(O(D(j))\times G^{n-1}$ and $O(D(n))\times G$ implies that;

$(\bigotimes_{j=1}^{n}(O(D(j))\times G^{n}$ is isomorphic to $(O(D(1)\times G)\otimes.....\otimes O(D(n)\times G)$

Hence by induction this expression must hold for all values of *n*. As an example let us we partition the event horizon *D* of a Black Hole as a subset of spacetime into *n* subsets *D(1), D(2),….,D(n),* then it follows, assuming that we have a section algebra *O(D* that in each subset we can locate a non-trivial section algebra *O(D(j))*. Choose a fibre algebra from each section algebra with a non-trivial state space which is generated by its extreme points – the pure states. This implies that corresponding to each partition element *D(j)* we can associate a pure state *f(j)*.

In quantum ergodic theory, the partition of a space into measurable subsets corresponds to a level of information about location in the space and this is measured by the information entropy of the partition [3]. Hawking's analysis of black hole dynamics; see for example [6] links the black hole event horizon surface area, partitions of that area, and measures of information entropy. The quantum equivalent of a partition is the density operator $\rho = \sum_k p_k y_k >< y_k = \sum_k p_k E_k$ ie the weighted sum of the projections $E_k$ onto the vector spaces $E_k H = [| y_k >]$ correponding to the normal pure state $\omega(y_k): A \to \langle y_k, Ay_k \rangle$. Then the von Neumann (quantum) entropy is defined as $-\sum_j p_j \log p_j$. We interpret it as an inverse measure of the amount of information that the quantum system in a given state will yield through measurement. The larger the entropy of the quantum system, the less information can be extracted.

By the 'no hair' theorem [7] each *n*-partition must have the same weighting $p_j = \frac{1}{n}$. The von Neumann entropy of this partition is thus given by $\log n$ ; proportional to its surface area *S*, at the micro-level when it is partitioned into *n* cells in the Planck regime. This relates to the classical macro- level Clausius definition of entropy *E*; $\oint \frac{\delta Q}{T} = \Delta E$ where, integrated over a Carnot cycle, a change in energy $\delta Q$ results in a change in entropy corresponding to a change in space-time curvature. [8].

**Measurability and Continuity**

The weak topology $\sigma(A(x)_*, A(x))$ can be defined on the predual $A(x)_*$ as the coarsest topology for which elements of the predual are continuous. It is defined by a set of semi-norms $p = |f|$ for $f$ a density matrix linear functional which as a set are separating for $A(x)_*$. Making minimal assumptions we let $\alpha : g \to \alpha_g$ be a weakly measurable representation of the compact Lie group $G$ as automorphisms of **A(x)**. By this we mean that the induced mapping[1] $\nu : g \to f \circ \alpha_g^{-1} : G \to \mathbf{A(x)}_*$ is measurable for Haar measure on $G$ and the $\sigma(A(x)_*, A(x))$ topology on $A(x)_*$. Since every positive element of $\mathbf{A(x)}_*$ is a countable sum of vector states this is equivalent to the definition that $\nu : g \to \omega_x \circ \alpha_g : G \to \mathbf{A(x)}_*$ is measurable for all $x$ in the fibre Hilbert space $F(x)$.

Given that the induced mapping $\nu : g \to f \circ \alpha_g : G \to \mathbf{A(x)}_*$ is measurable in the sense now defined above we have from [9];

$$\|\nu(g) \circ f(A) - \nu(h) \circ f(A)\| \leq \|\nu(g) \circ f - \nu(h) \circ f\| \|A\| \to 0 \text{ as } g \to h$$

This demonstrates the following result, which allows the extension of continuous gauge automorphic representations of compact Lie groups to their tensor product such as the Standard Model gauge group $SU(3) \otimes SU(2) \otimes S(1)$;

*For the induced representation $\nu : g \to f \circ \alpha_g : G \to \mathbf{A(x)}_*$ on the predual of A(x), weak measurability is equivalent to weak continuity.*

We next show, as we did for local diffeomorphism-invariant quantum states [1, 4] that quantum states invariant under the action now of compact Lie groups are common in the sense that the weakly closed convex hull of every normal state contains such a state. We are now dealing with groups such as *SU(n)* which are both compact and non-abelian thus different techniques are required. To achieve this result we have developed a new idea based

---

[1] For the rest of the paper we will ignore the inverse symbol in his definition to ease notational clutter.

on group stabiliser theory which we call Wigner sets. These are complementary to little groups.

**Wigner Sets and the Finite Intersection Property**

Given a density matrix quantum state $f$, and a weakly measurable representation $g \to \alpha_g$ of a compact Lie group $G$ as gauge automorphisms of the fibre algebra $\mathbf{A(x)}$; define the closed convex hull; $X(f) = \overline{co}\{f \circ \alpha_g ; g \in G\} \subset \mathbf{A(x)}_*$ with closure in the $\sigma(A(x)_*, A(x))$-topology. Define the group of isometric and $\sigma(A(x)_*, A(x))$-continuous transformations mapping $X(f) \to X(f)$ by $v(G) = \{v(g) : x \to x \circ \alpha_g ; g \in G, x \in X(f)\}$.

Mathematically, we note that since $G$ is compact and $f \circ \alpha : G \to \mathbf{A(x)}_*$ is weakly measurable and thus weakly continuous; this implies that $f \circ \alpha(G)$ is $\sigma(A(x)_*, A(x))$-compact. The Krein-Smulian theorem [10], then shows that $X(f)$ is also a $\sigma(A(x)_*, A(x))$-compact set. Thus $X(f)$ is a non-void $\sigma(A(x)_*, A(x))$-compact convex subset of the locally convex Hausdorff linear topological space of ultraweakly continuous linear functionals acting on the fibre algebra $\mathbf{A(x)}$. The group of mappings $v(G) = \{v(g) : x \to x \circ \alpha_g ; g \in G, x \in X(f)\}$ is, as we will now show, a non-contracting (semi)-group of weakly continuous affine maps of $X(f)$ onto itself.

If $x$, $y$ are in $X(f)$ then;
$\|v(g)x - v(g)y\| = \sup_{\|A\|\leq 1} \|v(g)x(A) - v(g)y(A)\| = \sup_{\|A\|\leq 1} \|x \circ \alpha_g(A) - y \circ \alpha_g(A)\| = \|x - y\|$
since each $\alpha_g$ is a continuous bijection. Thus if $x \neq y$ and
$S = $ the weak closure of $v(g)x(A) - v(g)y(A); A \in \mathbf{A(x)}$ then $0 \notin S$.

Thus each mapping $v(g)$ is non-contracting. Also, each $v(g)$ is affine, since for $A$ in $\mathbf{A(x)}$ and $0 \leq \lambda \leq 1$.

$$v(g) \circ \lambda x + (1-\lambda) y \ (A) = \lambda x(\alpha_g(A)) + (1-\lambda) y(\alpha_g(A)) = \lambda v(g) \circ x \ A + (1-\lambda) v(g) \circ y \ A$$

We can therefore, apply the Ryll-Nardzewski fixed point theorem [11] to establish the existence of an invariant normal state contained in $X(f)$.

The physical implications highlight the role of what we have termed Wigner sets.

Given $g \in G$, define the Wigner set of the mapping $\nu(g): X(f) \to X(f)$ as the stabiliser set;
$\mathcal{F}(\nu(g)) = \{x \in X(f); \nu(g) \circ x = x\}$.

More generally, given a finite subset $\{g(j) \in G; j =, 2,...n\}$ and corresponding mappings $\{\nu(g(j)); j =, 2,...n\}$, we can construct the affine mapping $\frac{1}{n}\left(\sum_j \nu(g(j))\right): X(f) \to X(f)$.

We then define $\mathcal{F}\{\nu(g(j)); j =, 2,...n\} \triangleq \mathcal{F}\left(\frac{1}{n}\left(\sum_j \nu(g(j))\right)\right)$

We assert that the following relationship between Wigner sets applies;

$$\bigcap_j \mathcal{F}(\nu(g(j))) = \mathcal{F}\left(\frac{1}{n}\left(\sum_j \nu(g(j))\right)\right) \quad \text{...........................(1)}$$

Clearly if $x \in \bigcap_j \mathcal{F}(\nu(g(j))) \Rightarrow \nu(g(j)) \circ x = x \; \forall \; j = 1,..,n \Rightarrow \frac{1}{n}\left(\sum_j \nu(g(j))\right) \circ x = x$

$\Rightarrow x \in \mathcal{F}\left(\frac{1}{n}\left(\sum_j \nu(g(j))\right)\right)$ therefore $\bigcap_j \mathcal{F}(\nu(g(j))) \subseteq \mathcal{F}\left(\frac{1}{n}\left(\sum_j \nu(g(j))\right)\right)$

To complete the proof of equation (1) we now need to show that;

$$\mathcal{F}\left(\frac{1}{n}\left(\sum_j \nu(g(j))\right)\right) \subseteq \bigcap_j \mathcal{F}(\nu(g(j))) \quad \text{..........................(2)}$$

This relation is trivially true for n=1.

If the result (2) is false then;

There is a minimum positive integer $r \geq 2$ for which (2) fails……………………………..(3)

For ease of exposition, denote the mapping $T(r) = \left( \dfrac{1}{r} \sum_{j=1,...,r} \nu(g(j)) \right)$ and the mappings

$T_j = \nu \; g \; j \; , j=1,......,n.$ so that $T(r) = \dfrac{1}{r} \left( \sum_{j=1,...,r} T_j \right)$ all acting on $X(f)$.

As noted earlier, since $G$ is compact, the Krein-Smulian theorem, [10] shows that $X(f)$ is a $\sigma(A(x)_*, A(x))$- compact set. $T(r)$ is an affine mapping on the compact convex set $X(f)$ so by Schauder's extension of the Brouwer fixed point theorem [10] has a fixed point $x(r)$ in $X(f)$. This implies that;

$$T(r) \circ x(r) = \dfrac{1}{r} \left( \sum_{j=1,...,r} T_j \circ x(r) \right) = x(r).$$

By the definition of $r$ as the minimum integer for which our assertion fails;

we must have an integer $k \leq r$ with $T_k \circ x(r) \neq x(r)$ ……………………………….(4)

Suppose now that $T_r \circ x(r) = x(r)$

Then

$rx(r) = rT(r) \circ x(r) = \left( \sum_{j=1,...,r} T_j \circ x(r) \right) = (T_1 + ... + T_{r-1}) \circ x(r) + T_r x(r) = (T_1 + ... + T_{r-1}) \circ x(r) + x(r)$

$\Rightarrow x(r) = \dfrac{1}{(r-1)} (T_1 + ... + T_{r-1}) \circ x(r) \Rightarrow x(r) \in \mathcal{F}(T(r-1)) = \bigcap_{j=1,...,r-1} \mathcal{F}(\nu(g(j)))$

$\Rightarrow T_j x(r) = x(r)$ for all $j = 1..., r$

But by definition of $k$ at (4) we have a logical contradiction.

By reordering the sum $\left( \sum_{j=1,...,r} T_j \circ x(r) \right)$ we can relabel $r$ by any $j=1,...,r.$.

In summary, we have shown that if our assertion at equation (2) is false then;

$T_j \circ x(r) \neq x(r)$ for all $j = (1,...,r)$.................................................(5)

Let $\mathcal{G}$ be the subgroup generated by the finite set $T_j = \nu(g_j); j = (1,...,r)$ and let $K(r)$ be the weakly closed convex hull of $\{Sx(r); S \in \mathcal{G}\}$. Since the set of mappings $\nu G$ is non-contracting, and $T_j x(r) \neq x(r)$ for all $j = (1,...,r)$ there is a continuous semi-norm $p$ on $X(f)$ with;

$$p(ST_j x(r) - Sx(r)) > \varepsilon \; \forall j = 1,...,r \;, S \in \mathcal{G} \quad \text{................................................(6)}$$

We will prove that this implication of (3) also gives rise to a contradiction, exploiting now part of the argument in [12].

Since $\mathcal{G}$ is finitely generated, it follows that $K(r)$ is separable, as well as being a weakly compact, convex subset of the dual space of **A(x).** It has the appropriate geometric and topological properties for [12] to apply. Thus for the given $\varepsilon > 0$ above there is a proper closed convex subset $C$ of $K(r)$ with ;

$$\text{the } p\text{-diameter of } K(r) \setminus C = \sup\{p(x - y); x, y \in K(r) \setminus C\} \leq \varepsilon.$$

$K(r) \setminus C$ is non-void, and $K(r)$ is the weakly closed convex hull of $\{Sx(r); S \in \mathcal{G}\}$, thus we can choose a mapping $S$ such that $Sx(r) \in K(r) \setminus C$. Now our chosen $S$ is an affine mapping, therefore;

$$Sx(r) = ST(r)x(r) = \frac{1}{r}\left(\sum_j ST_j x(r)\right)$$

$C$ is a convex set, thus not every $ST_j x(r)$ can be a member of $C$, otherwise the expression above would then imply that $Sx(r) \in C$. Thus, for some $j$, $ST_j x(r) \in K(r) \setminus C$.

We thus have;

$Sx(r) \in K(r) \setminus C$ and $ST_j x(r) \in K(r) \setminus C$. But for the seminorm $p$, $p - diam(K(r) \setminus C) \leq \varepsilon$. Thus for some $j$ ;

$$p(ST_j x(r) - Sx(r)) \leq \varepsilon.$$

This contradicts (6), showing that there is no minimum positive integer $r \geq 2$ for which (2) fails. and we have proved our assertion on the relation between Wigner sets;

$$\bigcap_j \mathcal{F}(\nu(g(j))) = \mathcal{F}\left(\frac{1}{n}\sum_j \nu(g(j))\right)$$

**Invariant Normal States**

It is now easy to prove that $X(f)$ contains a fixed point for the group of isometric and $\sigma(A(x)_*, A(x))$-continuous transformations $\nu(G) = \{\nu(g)(x) = x \circ \alpha_g ; g \in G, x \in X(f)\}$.

We have that $\left(\frac{1}{n}\sum_j \nu(g(j))\right)$ is a $\sigma(A(x)_*, A(x))$- continuous affine mapping on the compact convex set $X(f)$ it has a fixed point $x$ (applying again Schauder's fixed point theorem [10]). Then $x \in \mathcal{F}\left(\frac{1}{n}\sum_j \nu(g(j))\right)$.

The expression;

$$\bigcap_j \mathcal{F}(\nu(g(j))) = \mathcal{F}\left(\frac{1}{n}\sum_j \nu(g(j))\right)$$

Shows that the Wigner sets $\mathcal{F}(\nu(g))$ have the finite intersection property since $\bigcap_j \mathcal{F}(\nu(g(j)))$ is non-void. Clearly each Wigner set is a $(\sigma(A(x)_*, A(x)))$ closed subset of the compact set $X(f)$ thus $\bigcap_g \mathcal{F}(\nu(g)) \neq \emptyset$. If $h \in \bigcap_g \mathcal{F}(\nu(g)) \Rightarrow h = \nu_g \circ h = h \circ \alpha_g \forall g \in G$.

Thus $h$ is the required invariant quantum state.

*The proof shows that quantum states invariant under the action of compact Lie groups are common in the sense that the weakly closed convex hull of every normal state contains such an invariant state.*

**Separating States and Locally Invariant Fields**

As in [4] with possible applications to weak measurements as in [13-15] we define a *separating (or total) family* of $G$-invariant quantum states to be a finite or countable subset of the state space such that the positive kernel of $S$ is zero. Applied to a single state this implies that the GNS mapping $\pi(f)$ is an isomorphism, and given an observable $A$ in the positive subset of the fibre algebra $\mathbf{A}(\mathbf{x})$, $f(A^*A) = 0 \Rightarrow A = 0$. This implies that if $f = \omega_\xi \circ \pi(f)$ then our definition is a weaker form than the definition of $\xi$ as a separating vector for the algebra $\pi(f)$ (**A(x)**) ; see for example [16]. If the Hilbert space *F(x)* on which the fibre algebra **A(x)** acts is separable then by the argument of [4] there is a separating (in our sense) normal state $\varphi$ on **A(x)**. and each element of the weakly closed convex hull $X(f) = \overline{co}\{f \bullet \alpha_g ; g \in G\}$ is also separating. Thus there is a separating normal invariant quantum state $\Psi = \Psi\ A(x)$ for each fibre algebra.

*Stitching these together in the obvious way we can create a locally G- invariant quantum state $\Psi$ of the field $A = A\ x\ ; x \in D$ with $\Psi|_{\mathbf{A\ x}} = \Psi\ A(x)$.*

**References**

1. J Moffat, Oniga T, Wang C 'Unitary Representations of the Translational Group Acting as Local Diffeomorphisms of Space-Time' J Phys Math 8:2, DOI: 10.4172/2090-0902.1000233 (2017).

2. J Moffat, Oniga T, Wang C 'A New Approach to the Quantisation of Paths in Space-Time' J Phys Math, 8:2, DOI: 10.4172/2090-0902.1000232. (2017).

3. J Moffat, Oniga T, Wang C 'Ergodic Theory and the Structure of Non-Commutative Space-Time.' J Phys Math 8.2,. DOI:10.4172/2090-0902.1000229 (2017).

4. J Moffat, Wang C 'Factorial Unitary Representations of the Translational Group, Invariant Pure States and the Supersymmetric Graviton' Arxiv [math-ph]:1708.03802 submitted to J Phys Maths (Oct 2017).

5. J Moffat 'On Groups of Automorphisms of the Tensor Product of von Neumann Algebras' Math Scand 34 (1974).

6. R Wald 'Quantum Field Theory in Curved Space-Time and Black Hole Thermodynamics' Univ of Chicago Press 1994

7. Ashtekar A 'The Simplicity of Black Holes' ; Physics 8, 34 (2015)

8. T Jacobson 'Thermodynamics of Spacetime: The Einstein Equation of State' Phys. Rev. Lett. 75 (1995).

9. J Moffat 'Continuity of Automorphic Representations' Math Proc Camb Phil Soc 74, (1973).

10. N. Dunford and J. Schwartz, Linear operators. Vol I, Interscience, New York, 1958.

11.     C Ryll-Nardzewski  'On Fixed Points of Semi-Groups of Endomorphisms of Linear Spaces'. Proc. 5th Berkeley Symp. Probab. Math. Stat. Univ. California Press 2(1) (1967).

12.     I  Namioka, E. Asplund  'A Geometric Proof of Ryll-Nardzewski's Fixed Point Theorem.' Bull. Amer. Math. Soc. 73 (1967)

13.     T Denkmayr., Geppert, H: et al  'Observation of a quantum Cheshire Cat in a matter-wave interferometer experiment'. Nature Communications, DOI: 10.1038/ncomms5492 (29 Jul 2014).

14.      A Matzkin, Pan, A. K  'Three-box paradox and Cheshire cat grin: the case of spin-1 atoms'. Journal of Physics A: Mathematical and Theoretical 46, 315307 (2013);

15.      Y Aharonov., Popescu, S et al: 'Quantum Cheshire Cats'. New Journal of Physics 15, 1130a15 (2013).

16.     H Araki 'Mathematical Theory of Quantum Fields'  Oxford University Press ppbk (2009)